\documentclass[conference]{IEEEtran}
\IEEEoverridecommandlockouts
\usepackage{cite}
\usepackage{algpseudocode}
\usepackage{algorithmicx,algorithm}
\usepackage{epstopdf}
\usepackage{graphicx}
\usepackage{amsmath,amssymb,amsfonts}
\usepackage{bm}
\usepackage{subcaption}
\usepackage{textcomp}
\usepackage{xcolor}
\usepackage{tikz}
\usepackage{dsfont}
\usetikzlibrary{shapes.geometric, arrows,arrows.meta}
\def\BibTeX{{\rm B\kern-.05em{\sc i\kern-.025em b}\kern-.08em
    T\kern-.1667em\lower.7ex\hbox{E}\kern-.125emX}}
\allowdisplaybreaks
\begin{document}
\title{Mean-Field Approximation based Scheduling for Broadcast Channels with Massive Receivers}
\author{\IEEEauthorblockN{Changkun Li*$\dagger$,~\IEEEmembership{Graduate Student Member, IEEE}, Wei Chen*$\dagger$, \emph{Senior Member, IEEE},\\~and~Khaled B. Letaief$\ddagger$,~\IEEEmembership{Fellow,~IEEE}}
\IEEEauthorblockA{* Department of Electronic Engineering, Tsinghua University, Beijing, CHINA, 100084\\
 $\dagger$ Beijing National Research Center for Information Science and Technology (BNRist)\\
 $\ddagger$ School of Engineering, Hong Kong University of Science and Technology, Hong Kong\\
 Email: lck19@mails.tsinghua.edu.cn, wchen@tsinghua.edu.cn, eekhaled@ust.hk}
  \thanks{This work is
   supported in part by the National Key R$\&$D Program of China under Grant 2018YFB1801102, the National Natural Science Foundation of China under grant No. 61971264, and Beijing Natural Science Foundation under grant No. 4191001.} 
}
\maketitle

\begin{abstract}
The emerging Industrial Internet of Things (IIoT) is driving an ever increasing demand for providing low latency services to massive devices over wireless channels. As a result, how to assure the quality-of-service (QoS) for a large amount of mobile users is becoming a challenging issue in the envisioned sixth-generation (6G) network. In such networks, the delay-optimal wireless access will require a joint channel and queue aware scheduling, whose complexity increases exponentially with the number of users. In this paper, we adopt the mean field approximation to conceive a buffer-aware multi-user diversity or opportunistic access protocol, which serves all backlogged packets of a user if its channel gain is beyond a threshold. A theoretical analysis and numerical results will demonstrate that not only the cross-layer scheduling policy is of low complexity but is also asymptotically optimal for a huge number of devices.
\end{abstract}

\section{Introduction}
With the development of the fifth-generation (5G) and future sixth-generation (6G) \cite{6G} wireless networks, the Industrial Internet of Things (IIoT) has emerged as an important class of applications. The emerging IIoT applications stimulated an ever-increasing demand to provide low latency services for massive devices \cite{mURLLC}. However, the scarce spectrum resources and limited transmission power is a major challenge and is hindering the support of required quality of service (QoS) for massive users. How to efficiently utilize the limited resources to support QoS for massive users is becoming critical and has been gaining much attention in recent years.

In order to reduce the communication latency, a line of works focused on the design of efficient stochastic optimization based resources allocation schemes. In \cite{Dynamic_ser}, a dynamic server allocation scheme for parallel queues with random varying connectivity was investigated. It was shown that the allocation policy which serves the longest queue first performs well in stabilizing the system. In \cite{Berry_queue}, an optimal delay-power tradeoff was revealed in the case of asymptotically small delays with fading channels. In \cite{Chen_queue}, a unified framework for queue-based transmission scheduling was proposed. The optimal scheduling policy was proved to have a threshold-based structure. Work \cite{Chen_queue} has been extended to different scenarios by works \cite{Liu, Zhao,NOMA}. In \cite{Liu} and \cite{Zhao}, the optimal queue-aware scheduling policies under the multi-state fading channels and the Markov arriving process were investigated, respectively. In \cite{NOMA}, a queue status based non-orthogonal multiple access was investigated to reduce the latency in multiple user access scenarios. This line of works focused on the queue status based resource allocation optimization to match the traffic dynamics of users, thereby achieving better performance than other allocation schemes. However, the complexity of this line of works increases exponentially with the number of users, leading to an intrinsic difficulty in analyzing and optimizing the queue status based resource allocation scheme. Motivated by this, this paper aims to apply a mean-field approximation method to analyze the queue status based resource allocation schemes with massive users.

In this paper, we consider a downlink wireless communication scenario, which includes a base station (BS) and a massive number of users where the BS transmits each user's data packets through a common broadcast channel. To serve massive users with low implementation complexity, the BS applies the time division multiple access (TDMA) protocol. In the order to minimize the average total queueing latency, we derive an optimal channel resources allocation policy by analyzing established stochastic optimization problems. Under the optimal allocation policy, the BS serves the user with the best channel state first. Moreover, the optimal policy is based on the channel status and queue status of all the users. As a result, it is not a trivial work to exactly evaluate the optimal policy. To address this challenge, we propose a mean-field approximation method to approximately evaluate the performance of the optimal policy. When the user number becomes massive, all users' influences on a certain user's queue dynamic characteristics become certain. As a result, we can decompose the influences among each user. After decomposing the complex massive user system, we establish an equivalent one-dimensional single user queue model for each user to analyze the queueing latency. Numerical results will demonstrate that the proposed mean-field approximation method can well estimate the optimal policy's delay performance.

\section{System Model}
\begin{figure}[t]
    \centerline{\includegraphics[width=8cm]{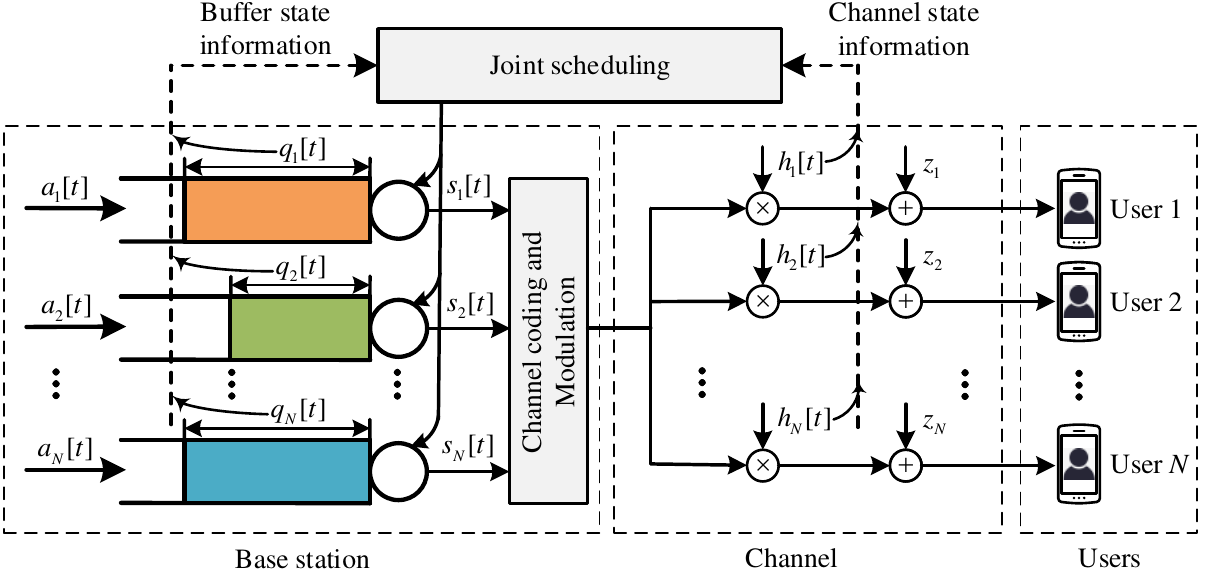}}
    \caption{The proposed system model.}
    \label{sysModel}
\end{figure}

As depicted in Fig. \ref{sysModel}, we consider a multi-user downlink wireless communication scenario, which is comprised of a BS and $N(\gg1)$ users. The transmission power at the BS is set to a constant $\mathcal{P}$ to help reduce implementation complexity. At the BS, each user is allocated with a separate buffer to store the packets of independent identically distributed (i.i.d.) arrival processes. To serve the packets backlogged in the buffers, all users share a common broadcast channel to transmit their data packets. Moreover, a central scheduler is adopted to allocate the common channel resources based on the buffer states and channel states of each user.

\subsection{The Physical Layer}
In this subsection, we present the physical layer of the downlink wireless system. In this downlink wireless system, time is denoted by $\tau$. We divide time into slots. Each timeslot has a duration of $T$ seconds. The $t$th timeslot denotes the time interval $[tT,(t+1)T]$. Moreover, the bandwidth of the channel is denoted by $W$. We consider a block fading channel model, which has been used to model the slowly varying flat-fading channel. Assume each user's channel quality follows an i.i.d fading process with an i.i.d additive white Gaussian noise (AWGN). The channel gain of each user in each timeslot over a bandwidth $W$ does not change much. Let $h_n[t]$ and $\sigma^2$ denote the magnitude of channel gain of user $n$ in timeslot $t$ and the noise power at each user's receiver, respectively.

To serve all the users through a common broadcast channel, a general superposition coding scheme can achieve good performance but with intolerably high complexity in analysis and implementation when $N\gg 1$. To avoid tackling the high complexity of the superposition coding scheme, we apply a suboptimal time division multiple access (TDMA) protocol. Especially, it was shown in \cite{TDMA} that the TDMA achievable capacity region converges to the capacity region achieved by the superposition coding scheme as the power decreases. Under the TDMA protocol, the BS shall allocate all the transmission power $\mathcal{P}$ and bandwidth $W$ to one user at a time instant $\tau$. Let $\mathds{1}_n(\tau)\in\{0,1\}$ denote a channel allocation policy as a time instant $\tau$. Especially, $\mathds{1}_n(\tau)=1$ indicates that user $n$ is allowed to access the common channel at the time instant $\tau$. If $\mathds{1}_n(\tau)=0$, the user $n$ is not served at time instant $\tau$. In particular, we have $\sum_{n=1}^{N}\mathds{1}_n(\tau)\le 1$. Let $R_n(\tau)$ denote the information rate of user $n$ at time instant $\tau$. According to the channel capacity \cite{IT}, we have
\begin{equation}\label{capacity}
R_{n}(\tau) \le \log(1+h_n^2[t]\mathds{1}_n(\tau)\rho),
\end{equation}
where $\rho=\frac{\mathcal{P}}{\sigma^2}$ denotes the signal-to-noise ratio.

\subsection{Discrete-time Queue Model}
We consider a discrete-time continuous-length queue model, which can be obtained by sampling the fluid model with sampling rate $1/T$. Next, we present the variation of user $n$'s queue length in a timeslot. Assume that the arrival process of each user is i.i.d.. At the start of timeslot $t$, $a_n[t]\in[A]$\footnote{In this paper, we denote by $[A]$ the set $\{0,1,...,A\}$.} packets arrive at user $n$'s buffer. Let the probability $\Pr\{a_n[t]=a\}$ be denoted by $\theta_a$, where $a\in[A]$ and $\sum_{a=0}^A\theta_a=1$. Without loss of generality, suppose that all packets are the same in size and each packet contains $L$ bits. As a result, the average number of bits arriving at each user's buffer in a timeslot is given by 
\begin{equation}
\lambda = L\sum_{a=0}^A a\theta_a.
\end{equation}

Let $q_n[t]$ denote the number of bits remained in user $n$'s buffer by the end of the timeslot $t$. The number of bits in user $n$'s buffer that are transmitted in timeslot $t$ is denoted by $s_n[t]$. Since the number of bits in a user's buffer cannot be negative, $s_n[t]$ shall satisfy that $s_n[t]\le q_n[t-1]+a_n[t]L$. By the end of timeslot $t$, the number of bits remained in user $n$'s buffer, i.e. $q_n[t]$, is updated as
\begin{equation}\label{update}
q_n[t] = q_n[t-1]+a_n[t]L-s_n[t].
\end{equation}
Let $S[t]=\sum_{n=1}^Ns_n[t]$ denote the total number of bits transmitted in timeslot $t$. According to the Nyquist's theorem, the value of $S[t]$ is given by
\begin{equation}\label{St}
S[t]= W\int_{tT}^{(t+1)T}\sum_{n\in\mathcal{N}(\tau)}\log\bigg(1+h^2_{n}[t]\mathds{1}_{n}(\tau)\rho\bigg){\rm d}\tau.
\end{equation}

\section{Delay-Optimal Scheduling with Massive Users}
In this section, we aim to derive an optimal scheduling policy to minimize the average total queueing latency in the aforementioned downlink wireless system.

In each timeslot $t$, the BS dynamically adjust the power allocated to each user based on each user's queue state and channel state. A power allocation policy in timeslot $t$ can be given by $\{\mathds{1}_n(\tau),n\in\{1,2,...,N\},\tau\in[tT,(t+1)T\}$.Based on the power allocation policy, each user's queue length updates as Eq. (\ref{update}). According to little's law, we can formulate a following optimization problem to minimize the average total queueing delay, which is given by
\begin{subequations}\label{opt1}
\begin{align}
\min&\quad \frac{1}{\lambda}\sum_{i=1}^N\mathbb{E}\{q_n[t]\},\label{obj_1}\\
s.t.&\quad \sum_{n\in\mathcal{N}(\tau)}\mathds{1}_n(\tau)\le1,\forall\tau\ge 0,
\end{align}
\end{subequations}
where $\mathbb{E}\{\cdot\}$ denotes the expectation operator.

However, it is a non-trivial work to solve the optimization problem (\ref{opt1}). The intrinsic high complexity of optimization problem (\ref{opt1}) is due to we aim to minimize the average queueing delay in a infinite time horizon by determining the power allocated at each time instant $\tau$. To overcome the difficulty in solving the optimization problem (\ref{opt1}), we aim to convert Problem (\ref{opt1}) into an equivalent optimization problem, which is given in the following theorem.

\noindent\textbf{Theorem 1:} The optimization problem (\ref{opt1}) is is asymptotically equivalent to the following optimization problem as $N\to\infty$. 
\begin{subequations}\label{opt}
\begin{align}
\max&~S[t]\hspace{-0.1em}=\hspace{-0.1em}W\hspace{-0.3em}\int_{tT}^{(t+1)T}\hspace{-0.8em}\sum_{n\in\mathcal{N}(\tau)}\hspace{-0.3em}\log\bigg(1+h^2_{n}[t]\mathds{1}_{n}(\tau)\rho\bigg){\rm d}\tau,\label{opt_St}\\
s.t.&\sum_{n\in\mathcal{N}(\tau)}\mathds{1}_n(\tau)\le1,\forall\tau\in[tT,(t+1)]. 
\end{align}
\end{subequations}
\begin{IEEEproof}
Due to space limitations, we mainly present the main idea of the proof. We first present an equivalent expression of the objective function Eq. (\ref{obj_1}). Let $Q[t]=\sum_{n=1}^{N}(q_n[t-1]+a_n[t]L)$ denote the total number of bit that can be transmitted in timeslot $t$.
According to Eq. (\ref{update}), Eq. (\ref{obj_1}) is equivalent to
\begin{equation}\label{eqv}
\begin{split}
&\frac{1}{\lambda}\sum_{n=1}^N\mathbb{E}\{q_n[t]\}=\frac{1}{\lambda}\sum_{n=1}^N\mathbb{E}\{q_n[t-1]\}\\
&=\frac{1}{\lambda}\mathbb{E}\{Q[t]-\sum_{n=1}^Na_n[t]L\}=\frac{1}{\lambda}\mathbb{E}\{Q[t]\}-N.
\end{split}
\end{equation}
As a result, we shall minimize $\mathbb{E}\{Q[t]\}$. For a policy that does not maximize $S[t]$ in some timeslots, we prove that their exists a new policy can achieve lower or equal $Q[t]$ for all $t$. The main idea of the proof is completed.
\end{IEEEproof}

Let the optimal channel allocation scheme at time $\tau$ be denoted by $\{\mathds{1}^*_n(\tau)\}$, which is given in the following theorem. 

\noindent\textbf{Theorem 2:} One of the optimal channel allocation scheme at time $\tau$ is given by
\begin{equation}\label{con}
\mathds{1}^*_n(\tau)=\left\{
\begin{split}
&1,~ n = \arg\max_{n\in\mathcal{N}(\tau)}h_n[t],\\
&0, ~\text{otherwise}.
\end{split}
\right.
\end{equation}
\begin{IEEEproof}
We denote by 
\begin{equation}
\bar{R}(T_1)=\frac{1}{T_1}\int_{tT}^{tT+T_1}\sum_{n=1}^{|\mathcal{N}(\tau)|}\log_2\left(1+h_n^2[t]\mathds{1}_n(\tau)\rho\right){\rm d}\tau
\end{equation}
the average information rate over time interval $[tT,tT+T_1]$. It is obvious that we shall maximize $\bar{R}(T)$ to minimize (\ref{opt_St}). It can be verified that the solution $\{\mathds{1}^*_n(\tau)\}$ given by Theorem 2 can minimize $\bar{R}(T_1)$ for all $T_1\in[0,T]$. The proof is completed.
\end{IEEEproof}

According to Theorem 2, the BS shall serve each user from the user with the best channel state to the user with the worst channel state in each timeslot. The corresponding optimal scheduling policy is referred to as a good-channel-first-serve (GCFS) policy. 

Next, we derive $s_n[t]$ under the GCFS policy. Let $\{n_1,n_2,...,n_M\}$ denote the set of users whose buffer is not empty in timeslot $t$ just after the arrival process, i.e., $q_{n_m}[t-1]+a_{n_m}[t]L>0$ for $1\le m \le M$. The number $M$ is the number of users who need service in timeslot $t$. Without loss of generality, suppose that $h_{n_1}[t]\ge ... \ge h_{n_M}[t]$. Let 
\begin{equation}
v_{n_m}[t] = \frac{q_{n_{m}}[t-1]+a_{n_m}[t]L}{\log(1+h^2_{n_m}[t]\rho)}
\end{equation}
denote the number of channel symbols used to clear user $n_m$'s buffer. Under the GCFS policy, $s_n[t]$ is then given by the following corollary.

\noindent\textbf{Corollary 1:}
There exits a threshold $\iota$ given by
\begin{equation}
\iota = \mathop{\arg\max}_{m_t}\sum_{m=1}^{m_t}v_{n_m}[t] \le WT,1\le m_t\le M.
\end{equation}
For users in set $\{n_1,...,n_{\iota}\}$, $s_{n_m}[t]=q_{n_{m}}[t-1]+a_{n_m}[t]L$. For user $n_{\iota+1}$, 
\begin{equation}
s_{n_{\iota+1}}[t]=\Big(WT-\sum_{m=0}^{\iota}v_{n_m}[t] \Big)\log\Big(1+h^2_{n_{\iota+1}}[t]\rho\Big).
\end{equation}
Finally, for users in set $\{n_{\iota+2},...,n_{M}\}$, $s_{n_m}[t]=0$.
\begin{IEEEproof}
The proof is omitted due to space limitations.
\end{IEEEproof}
According to Corollary 1, the first $\iota$ users in set $\{n_1,...,n_{M}\}$ will surely get their data buffer cleared by the end of timeslot $t$.

\section{Mean-Field Approximation based\\ Delay Analysis}
In this section, we aim to analyze the delay performance of the GCFS policy. Under the GCFS policy, a discrete-time continuous-state $N$-dimensional Markov chain $\{q_n[t],1\le n\le N,t\ge 0\}$ is induced. 
Due to the curse of dimensionality, it is hard to analyze the performance of the GCFS policy by deriving the stationary distribution of the Markov chain when $N\gg 1$. To evaluate the delay performance of the GCFS policy, we apply a mean-field approximation method introduced in \cite{MFT_book}.

\subsection{Mean-Field Approximation and Threshold-based policy}
In this section, we aim to analyze the delay performance of the GCFS policy via a mean-field approximation method. Let $g(q_1,q_2,...,q_N)$ denote the stationary probability density function (PDF) of the Markov chain $\{q_n[t],1\le n\le N,t\ge 0\}$. According to the mean field approximation, we have
\begin{equation}\label{mft}
g(q_1,q_2,...,q_N)\approx\Pi_{n=1}^{N}g_n(q_n),
\end{equation}
when $N\to \infty$. This is due to that all users' impacts on a single user become certain when $N\to \infty$. In other words, the dependence among users vanishes in this downlink system as $N\to \infty$. Besides, since all users'  arrival process and channel quality are i.i.d., the PDFs $g_1(q_1),...,g_N(q_N)$ shall be identical. As a result, variables $q_n[t],1\le n\le N$ are i.i.d. when the induced Markov chain is stationary. According to the law of large numbers, we have
\begin{equation}\label{LLN}
\begin{split}
\lim_{N\to\infty}\frac{1}{N}\sum_{n=1}^{N}q_n[t]=\mathbb{E}\{q_n[t]\}={q}.
\end{split}
\end{equation}
Similarly, we have $\lim_{N\to\infty}\frac{1}{N}\sum_{n=1}^{N}a_n[t]L=\lambda$. That is to say that the average queue length and the average number of arriving bits in a timeslot are two constant when $N\to\infty$. The average number bits served in a timeslot, i.e., $\frac{1}{N}\sum_{n=0}^{N}s_n[t]$ shall also be constant that equals to $\lambda$. As a result, we have 
\begin{equation}\label{LLN1}
\lim_{N\to\infty}\frac{1}{N}\sum_{n=0}^{N}s_n[t]=\lambda.
\end{equation}

Under the GCFS policy, there exist a threshold $h_{\text{th}}$ on the channel gain that a user will get served when his channel gain is greater than $h_{\text{th}}$. Let $f(h)$ be the PDF of the channel gain $h_n[t]$. The cumulative distribution function of $h_n[t]$ is denoted by $F(x)$, where $F(x)=\int_{0}^{x}f(h)dh$. Set $G(x)=1-F(x)$. According to $h_{\text{th}}$, each user has the same probability $p=G(h_{\text{th}})$ to get served. As a result, there are about $Np$ users get served. Let $n_1,...,n_{Np}$ denote the $Np$ users where $h_{n_1}[t]\ge ...\ge h_{n_{Np}}[t]$. According to Corollary 1, the first $Np-1$ users will get their buffer cleared and the last user's bits are partially served. As a result, the total number of bits served in a timeslot is given by 
\begin{equation}
\sum_{n=0}^{N}s_n[t]=s_{Np}[t]+\sum_{m=1}^{Np-1}(q_{n_m}[t-1]+a_{n_m}[t]L),
\end{equation}
According to Eqs. (\ref{LLN}) and (\ref{LLN1}), we have 
\begin{equation}\label{LLN2}
\begin{split}
&\lim_{N\to\infty}\frac{1}{N}\Big(s_{Np}[t]+\sum_{m=1}^{Np-1}(q_{n_m}[t-1]+a_{n_m}[t]L)\Big)\\
=&\hspace{-0.3em}\lim_{N\to\infty}\frac{1}{N}\hspace{-0.3em}\sum_{m=1}^{Np-1}\hspace{-0.1em}(q_{n_m}[t-1]+a_{n_m}[t]L)\hspace{-0.1em}=\hspace{-0.1em}p({q}+\lambda)\hspace{-0.1em}=\hspace{-0.1em}\lambda.
\end{split}
\end{equation}
Eq. (\ref{LLN2}) indicates that $p$ and $h_{\text{th}}$ is a constant. As a result, a threshold-based policy which serves all backlogged packets of users whose channel gains are beyond $h_{\text{th}}$ can approximately achieves the minimum queueing latency, which has a lower complexity when compared with the GCFS policy. Besides, by Eq. (\ref{LLN2}), the average queue delay of each user is given by
\begin{equation}
D=\frac{{q}}{\lambda}=\frac{1}{p}-1.
\end{equation}

%
%
%
%
%

%

Next, we aim to establish a self-consistent equation to derive the unknown threshold $h_{\text{th}}$ and constant $p$. Based on the channel threshold $h_{\text{th}}$, the most average number of bits that can be transmitted in each timeslot can be computed as 
\begin{equation}
\Phi(h_{\text{th}})=\frac{\displaystyle WT\int_{h_{\text{th}}}^{\infty}f(h)log(1+h^2\rho){\rm d}h}{G(h_{\text{th}})}.
\end{equation}\
Let $h_u = \sup\{h|G(h)>0\}$. The feasible region of $h_{\text{th}}$ is given by $[0,h_u)$. The property of function $\Phi(h_{\text{th}})$ is presented in the following lemma. 

\noindent\textbf{Lemma 1:} $\Phi(h_{\text{th}})$ is non-decreasing in $h_{\text{th}}$.
\begin{IEEEproof}
We derive the derivative of the function $\Phi(h_{\text{th}})$, which is given by
\begin{equation}
\Phi'(\hspace{-0.1em}h_{\text{th}}\hspace{-0.1em})\hspace{-0.1em}=\hspace{-0.1em}\frac{\displaystyle WT\hspace{-0.1em}f(\hspace{-0.1em}h_{\text{th}}\hspace{-0.1em})\hspace{-0.3em}\int_{h_{\text{th}}}^{\infty}\hspace{-0.5em}f(\hspace{-0.1em}h\hspace{-0.1em})\hspace{-0.1em}\big(\hspace{-0.2em}\log(\hspace{-0.1em}1\hspace{-0.1em}+\hspace{-0.1em}h^2\rho\hspace{-0.1em})\hspace{-0.1em}-\hspace{-0.1em}\log(\hspace{-0.1em}1\hspace{-0.1em}+\hspace{-0.1em}h_{\text{th}}^2\rho\hspace{-0.1em})\hspace{-0.1em}\big){\rm d}h}{G^2(h_{\text{th}})}.
\end{equation}
Since 
\begin{equation}
\int_{h_{\text{th}}}^{\infty}\hspace{-0.5em}f(\hspace{-0.1em}h\hspace{-0.1em})\hspace{-0.1em}\big(\hspace{-0.2em}\log(\hspace{-0.1em}1\hspace{-0.1em}+\hspace{-0.1em}h^2\rho\hspace{-0.1em})\hspace{-0.1em}-\hspace{-0.1em}\log(\hspace{-0.1em}1\hspace{-0.1em}+\hspace{-0.1em}h_{\text{th}}^2\rho\hspace{-0.1em})\hspace{-0.1em}\big){\rm d}h\ge0,
\end{equation}
we have $\Phi'(\hspace{-0.1em}h_{\text{th}}\hspace{-0.1em})\ge 0$. As a result, $\Phi(h_{\text{th}})$ is a non-decreasing function of the threshold $h_{\text{th}}$.
\end{IEEEproof}
According to Lemma 1, the supremum of $\Phi(h_{\text{th}})$ is given by
\begin{equation}
\Phi_{\sup}=\lim_{h_{\text{th}}\to h_u}\Phi(h_{\text{th}})=WT\log(1+h_u^2\rho).
\end{equation}
The minimum of $\Phi(h_{\text{th}})$ is $\Phi(0)$.
In order to keep this queueing system stable, the average number of bits transmitted in each timeslot shall equal the average number of bits arriving in each timeslot. As a result, we have
\begin{equation}\label{self_consistent}
\Phi(h_{\text{th}}) = N\lambda.
\end{equation}
Eq. (\ref{self_consistent}) is the self-consistent equation we want to establish. By solving Eq. (\ref{self_consistent}), we can obtain the unknown threshold $h_{\text{th}}$ and the unknown constant $p$. In particular, when $\Phi_{\sup}< N\lambda$, this queueing system is unstable. More transmission power shall be applied in this case to enhance the service capability of the system. When $\Phi(0)>N\lambda$, we can reduce the transmission power to make full use of the spectrum resources. Since $\Phi(h_{\text{th}})$ is a non-decreasing function, we can apply Algorithm 1 to derive the threshold $h_{\text{th}}$. 

\begin{algorithm}[t]
\caption{The Algorithm to obtain the unknown channel quality threshold $h_{\text{th}}$.} 
\begin{algorithmic}[1]
\Require $N,\lambda,\rho,f(h),G(x),h_u,\epsilon$
\Ensure $h_{\text{th}}$
\If{$\Phi(0)\ge N\lambda$}
\State \textbf{return} 0
\EndIf
\If{$\Phi_{sup}\le N\lambda$}
\State \textbf{return} $h_u$
\EndIf
\State $h_l\gets0$, $h_r\gets h_u$, $h_{\text{th}}\gets \frac{1}{2}(h_l+h_r)$
\While {$|\Phi(h_{\text{th}})-N\lambda|>\epsilon$}
\If{$\Phi(h_{\text{th}})>N\lambda$}
\State $h_r = h_{\text{th}}$
\Else
\State $h_l = h_{\text{th}}$
\EndIf
\State $h_{\text{th}}\gets \frac{1}{2}(h_l+h_r)$
\EndWhile  
\end{algorithmic}
\end{algorithm}

\subsection{Equivalent Single User Queue via Decomposition}

In this subsection, we aim to approximate the stationary distribution of the each user's queue length based on the mean filed approximation results in the last subsection.  

\begin{figure}[t]
    \centerline{\includegraphics[width=8.4cm]{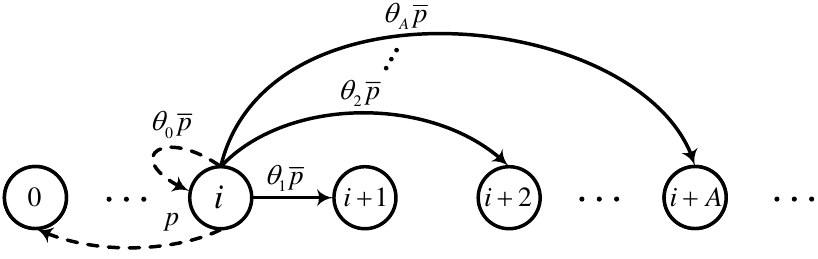}}
    \vspace{0.3em}
    \caption{The discrete-time discrete-state Markov chain model.}
    \label{Markov}
\end{figure}

Eq. (\ref{mft}) indicates that the dependence among users vanishes as $N\to \infty$. As a result, we can decompose the complex massive user downlink system into multiple equivalent single user systems. Eq. (\ref{LLN2}) indicates we can ignore the user whose bits are partially served. As a result, in each single user system, all packets in a user's buffer are approximately either all transmitted with probability $p$ or all remained in the buffer with probability $1-p$ in each timeslot. Let $\tilde{q}[t]$ denote the number of packets that remain in the buffer of user $n$ by the end of timeslot $t$. According to the arrival process $a_n[t]$, we can construct a discrete-time discrete-state Markov chain $\{\tilde{q}[t],t\ge 0\}$ to model the queue length dynamics of each single user system, which is illustrated in Fig. \ref{Markov}. Let $P_{i,j}=\Pr\{\tilde{q}[t+1]=j|\tilde{q}[t]=i\}$ be the one-step transition probability of this new Markov chain, which is given in the following lemma.

\noindent\textbf{Lemma 2:} The one-step transition probability of the new Markov chain is given by
\begin{equation}
P_{i,j}=\left\{
\begin{split}
&\theta_0+\bar{\theta}_0p, i=j=0,\\
&\theta_0\bar{p}, \qquad\; i=j>0,\\
&\theta_a\bar{p},\qquad\; i-j=a,0<a\le A,\\
&p,\qquad \quad\; i>0,j=0,\\
&0,\qquad \quad\;\text{otherwise},
\end{split}
\right.
\end{equation}
where $\bar{\theta}_0 = 1-\theta_0$ and $\bar{p} = 1-p$.
\begin{IEEEproof}
The proof is omitted due to space limitations.
\end{IEEEproof}

Let $\pi_i$ denote the steady-state probability of state $\tilde{q}[t]=i$ of the discrete-time discrete-state Markov chain $\{\tilde{q}[t],t\ge 0\}$. According to Lemma 2, we have
\begin{equation}\label{balance}
\sum_{j=i+A}^{\infty}\pi_{j}p = \sum_{j=0}^{A-1}\sum_{k=A-j}^{A}\pi_{i+j}\theta_k\bar{p},i\ge0.
\end{equation}
It should be noted that Eq. (\ref{balance}) is the local balance equation of the new discrete-time discrete-state Markov chain. Let $\chi_i=\sum_{j=i}^{\infty}\pi_{j}$. As a result, we have
\begin{equation}\label{diff}
\pi_{i} = \chi_i-\chi_{i+1}.
\end{equation}
By submitting Eq. (\ref{diff}) into Eq. (\ref{balance}), we have
\begin{equation}\label{diff_eq}
\chi_{i+A}c_A - \sum_{j=0}^{A-1}\chi_{i+j}c_j = 0,i\ge 0,
\end{equation}
where $c_A=p+\bar{\theta}_0\bar{p}$, and $c_j=\theta_{A-j}\bar{p}$ for $j\in[A-1]$. Eq. (\ref{diff_eq}) is a linear difference equation. To solve Eq. (\ref{diff_eq}), $A$ boundary conditions are needed, which are given in the following lemma.

\noindent\textbf{Lemma 3:} The $A$ boundary conditions of the linear difference Eq. (\ref{diff_eq}) are given by
\begin{equation}
\left\{
\begin{split}
&\chi_0 = 1,\\
&\chi_i = \frac{\displaystyle\chi_0\sum_{k=i}^{A}\theta_k\bar{p}+\sum_{j=1}^{i-1}\chi_j \theta_{i-j}\bar{p}}{p+\bar{\theta}_0\bar{p}},1\le i\le A-1.
\end{split}
\right.
\end{equation}
\begin{IEEEproof}
The proof is omitted due to space limitations.
\end{IEEEproof}

Next, we aim to solve the linear difference Eq. (\ref{diff_eq}) with $A$ boundary conditions by the unilateral $z$-transforms. Let $\tilde{\pi}(z)$ and $\tilde{\chi}(z)$ denote the unilateral $z$-transforms of sequences $\{\pi_i\}$ and $\{\chi_i\}$, respectively. The unilateral $z$-transform of Eq. (\ref{diff_eq}) is given in the following theorem.

\noindent\textbf{Theorem 3:} The unilateral $z$-transform of the linear difference Eq. (\ref{diff_eq}) is given by
\begin{equation}
\tilde{\chi}(z)=\frac{\displaystyle\sum_{j=0}^{A-1}\chi_j\bigg(\sum_{k=j+1}^{A-1}c_k z^{A-k} -c_Az^j\bigg)}{\displaystyle\sum_{j=0}^{A-1}c_j z^{A-j}-c_A}.
\end{equation}
\begin{IEEEproof}
The proof is omitted due to space limitations.
\end{IEEEproof}
Similar to Theorem 3, we can derive the unilateral $z$-transform of Eq. (\ref{diff}). By deriving the unilateral $z$-transform of Eq. (\ref{diff}), we can obtain
\begin{equation}\label{diff_z}
\tilde{\pi}(z) =\frac{(z-1)\tilde{\chi}(z) + 1}{z}. 
\end{equation}
We omit the derivation of Eq. (\ref{diff_z}) due to space limitations.
Let $\mathcal{Z}^{-1}(\cdot)$ denote the inverse unilateral $z$-transform operator. The steady-state probability of the discrete-state Markov chain is then given by
\begin{equation}\label{inverse}
\pi_i = \mathcal{Z}^{-1}(\tilde{\pi}(z)).
\end{equation}
In particular, when $A = 1$, the steady-state probabilities $\{\pi_i\}$ are given in the following corollary.

\noindent\textbf{Corollary 2:} When $A = 1$, the steady-state probabilities $\{\pi_i\}$ are given by
\begin{equation}
\begin{split}
&\pi_0 = \frac{p}{p\theta_0+\theta_1},\\
&\pi_i = \pi_0(1-\pi_0)^i, i\le 1.
\end{split}
\end{equation}
\begin{IEEEproof}
The proof is omitted due to space limitations.
\end{IEEEproof}

\section{Numerical Result}
In this section, numerical results are presented to demonstrate the theoretical analysis. In the simulations, the number of users and the maximum number of packets arriving at each user's buffer are set to $N=10^5$ and $A = 1$, respectively. The length of a packet is set to $L=1$. The bandwidth and the time interval of a slot are set to $W=1$ and $T=10^3$, respectively. The noise power at each user's receiver is set to $\sigma^2 = 10^{-12}$. Assume that the magnitude of user $n$'s channel gain, i.e., $h_n[t]$ is a Rayleigh random variable, which has a density distribution $f(h)=h e^{\frac{-h^2}{2}}$. In each timeslot, the packet arrival of each user's buffer follows a Bernoulli arrival process with parameter $\theta_1$. The BS allocates channel symbols according to Theorem 1. Each user's buffer evolves according to Eq. (\ref{update}) with each simulation running over $10^5$ timeslots. The simulation results and the results derived by the mean-field approximation are presented in the following figures. 

\begin{figure}[t]
    \centerline{\includegraphics[width=8.4cm]{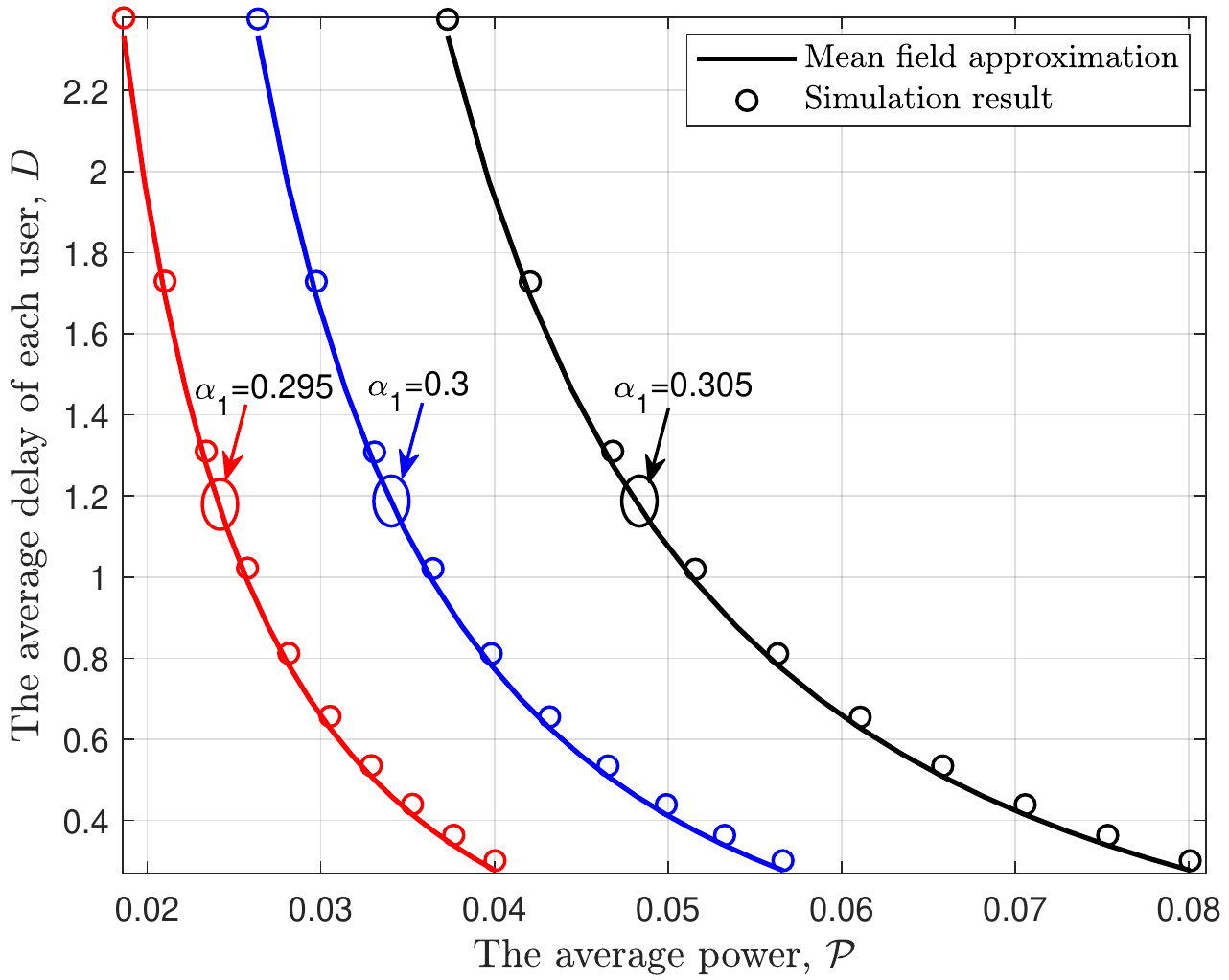}}
    \caption{The delay perfromance of GCFS policy.}
    \label{delay_performance}
\end{figure}

Fig. \ref{delay_performance} present the delay performance of the GCFS policy under different arrival rates $\theta_1$. In Fig. \ref{delay_performance}, the average packet delay approximated by the mean-field approximation method and the average bit delay derived through simulation are denoted by solid lines and marks `o', respectively. From Fig. \ref{delay_performance}, it can be seen that the average bit delay under the GCFS policy can be well approximated by the average packet delay derived through the mean-field approximation method proposed in this paper. Fig. \ref{delay_performance} also presents the basic tradeoffs between the queueing delay and the power consumption. When more transmission power is applied, the queueing delay decreases. 

\begin{figure}[t]
    \centerline{\includegraphics[width=8.4cm]{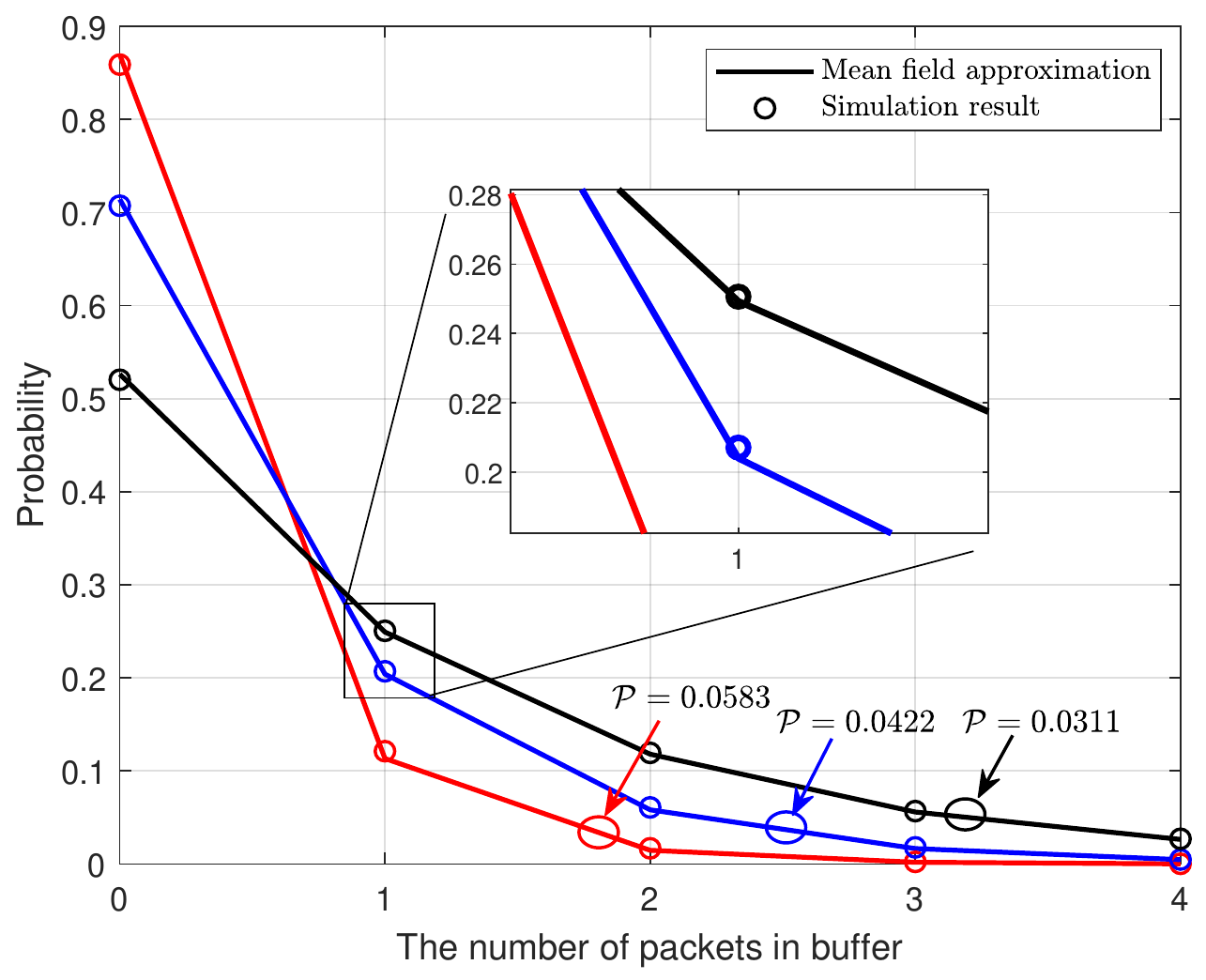}}
    \caption{The probability distribution of the number of packets in the buffer.}
    \label{dis_app}
\end{figure}

Fig. \ref{dis_app} present the probability distribution of the number of packets in one user's buffer under different transmission powers $\mathcal{P}$. In Fig. \ref{dis_app}, the arrival rate and the time interval of a slot are set to $\theta_1 = 0.6$ and $T=2\times10^3$, respectively. The probability distributions of the number of packets in one user's buffer derived by mean-field approximation and simulation are denoted by solid lines and marks `o', respectively. From Fig. \ref{dis_app}, it can be seen that the probability distributions of the number of packets in one user's buffer derived by the mean-field approximation and simulation match well. This demonstrates that the established discrete-state Markov chain depicted in Fig. \ref{Markov} well reflects the dynamic characteristics of each user's queue length.

\section{Conclusion}
In this paper, we have investigated the problem of queueing delay minimization in a downlink communication scenario with massive receivers. To minimize the average total queueing delay, we have derived the optimal scheduling policy, referred to as the GCFS policy, by analyzing the established stochastic optimization problem. Furthermore, to overcome the difficulty in analyzing the performance of the GCFS policy due to the massive users, we have applied a mean-field approximation method. We have obtained a threshold-based policy with lower complexity which can approximately achieve minimum delay. Moreover, we have shown that the complex downlink system with massive users can be decomposed into massive single-user scenarios. Besides, we have established a Markov chain model for the equivalent single-user scenario to analyze the stationary distribution of each user's queue length. Numerical results have demonstrated that the average queueing delay under the GCFS policy can be well evaluated by the mean-field approximation method.

\bibliographystyle{IEEEtran}
\linespread{0.92}
\bibliography{GC_multiuser_Li_Chen_Letaief}

\end{document}